\documentclass[sigconf]{acmart}
\usepackage{multirow}
\usepackage{enumitem}
\usepackage{makecell}
\newcolumntype{?}{!{\vrule width 1.5pt}}

\AtBeginDocument{%
  \providecommand\BibTeX{{%
    \normalfont B\kern-0.5em{\scshape i\kern-0.25em b}\kern-0.8em\TeX}}}

\setcopyright{acmlicensed}
\copyrightyear{2024}
\acmYear{2024}
\setcopyright{acmlicensed}\acmConference[RecSys '24]{18th ACM Conference on Recommender Systems}{October 14--18, 2024}{Bari, Italy}
\acmBooktitle{18th ACM Conference on Recommender Systems (RecSys '24), October 14--18, 2024, Bari, Italy}
\acmDOI{10.1145/3640457.3688195}
\acmISBN{979-8-4007-0505-2/24/10}

\begin{document}

\title{Does It Look Sequential? An Analysis of Datasets for Evaluation of Sequential Recommendations}

\author{Anton Klenitskiy}
\email{antklen@gmail.com}
\orcid{0009-0005-8961-6921}
\affiliation{
  \institution{Sber AI Lab}
  \city{Moscow}
  \country{Russian Federation}
}

\author{Anna Volodkevich}
\email{volodkanna@yandex.ru}
\orcid{0009-0002-7958-0097}
\affiliation{
  \institution{Sber AI Lab}
  \city{Moscow}
  \country{Russian Federation}
}

\author{Anton Pembek}
\email{apembek@bk.ru}
\orcid{0009-0005-1757-3379}
\affiliation{
  \institution{Sber AI Lab, Lomonosov Moscow State University (MSU)}
  \city{Moscow}
  \country{Russian Federation}
}
% \affiliation{
%   \institution{Lomonosov Moscow State University (MSU)}
%   \city{Moscow}
%   \country{Russian Federation}
% }

\author{Alexey Vasilev}
\email{alexxl.vasilev@yandex.ru}
\orcid{0009-0007-1415-2004}
\affiliation{
  \institution{Sber AI Lab}
  \city{Moscow}
  \country{Russian Federation}
}

% \renewcommand{\shortauthors}{Klenitskiy et al.}

\begin{abstract}
  
Sequential recommender systems are an important and demanded area of research. Such systems aim to use the order of interactions in a user's history to predict future interactions. The premise is that the order of interactions and sequential patterns play an essential role. Therefore, it is crucial to use datasets that exhibit a sequential structure to evaluate sequential recommenders properly.

We apply several methods based on the random shuffling of the user's sequence of interactions to assess the strength of sequential structure across 15 datasets, frequently used for sequential recommender systems evaluation in recent research papers presented at top-tier conferences. As shuffling explicitly breaks sequential dependencies inherent in datasets, we estimate the strength of sequential patterns by comparing metrics for shuffled and original versions of the dataset. Our findings show that several popular datasets have a rather weak sequential structure.
\end{abstract}






\maketitle

\section{Introduction}
\label{sec:intro}

% srs research area
% srs evaluation issues
Sequential Recommender Systems (SRSs) have been widely used to model short-term user preferences and user behavior over time, detect interest drifts of individual users, or identify short-term popularity trends \cite{quadrana2018sequence}. The recent rapid development of the SRSs, which is illustrated by various novel architectures, such as \cite{kang2018self, sun2019bert4rec, li2020time, liu2021noninvasive, xie2022contrastive}, brought the performance and evaluation questions, including the SRSs performance revision \cite{petrov2022systematic, klenitskiy2023turning, betello2023investigating}, evaluation protocols analysis \cite{hidasi2023widespread, ji2023critical} and a choice of the datasets for SRSs evaluation \cite{hidasi2023widespread, woolridge2021sequence}.

% % srs models development and hype
% With the development of recurrent neural architectures and Transformer-based models, a new round of SRS research and development has begun and brought SRS models, such as SASRec\cite{kang2018self},  BERT4Rec\cite{sun2019bert4rec}, further extended to incorporate addition information, such as TiSASRec\cite{li2020time}, NOVA-BERT\cite{liu2021noninvasive}, CL4SRec\cite{xie2022contrastive}.

% dataset selection for SRS evaluation
% A significant number of scientific papers employ a limited number of datasets, e.g. 3-5, to perform a model evaluation (links with examples). The choice of datasets is often determined by the necessity of being aligned with the previous works to enhance reproducibility. Moreover, datasets are often chosen heuristically, which could influence the observations and/or conclusions obtained \cite{sun2020we, chin2022datasets}. 
% As was shown by \cite{chin2022datasets}, the data characteristics have a significant impact on the evaluation results and thus it is important to use diverse datasets for proper evaluation. 
% Not only the datasets diversity should be considered, but an alignment of the datasets with the recommendation task.

According to \cite{hidasi2023widespread}, one of the core evaluation issues is the dataset-task mismatch, as the sequential recommendations only make sense if the data has sequential patterns. Thus, the datasets for SRS evaluation should be analyzed to determine the presence of such patterns. To the best of our knowledge, there are no well-established criteria to assess how good the dataset is for SRS evaluation. In our work, we aim to highlight the importance of determining the strength of sequential patterns during dataset selection and propose criteria for such assessment.

We propose to use three approaches to analyze the strength of sequential patterns. These approaches are based on the assumption that the sequential patterns in the dataset will be broken if the interactions in user sequences are shuffled in random order. Thus, we can estimate the strength of the sequential structure by comparing chosen metrics acquired on shuffled and unshuffled versions of data. 

The first approach, based on identifying sequential rules, is simple and model-agnostic. Two others consist of training a sequential model (SASRec and GRU4Rec in our experiments) and evaluating how the performance changes in terms of  NDCG@k and HitRate@k and how recommendation lists differ in terms of top-K Jaccard score \cite{oh2022ranklist} when the model is applied on original and shuffled sequences. 

In short, the main contributions of this paper are:
\begin{itemize}[topsep=2pt]

\item We propose a set of three assessment approaches based on user sequences shuffling to evaluate the presence and strength of a dataset's sequential structure.
\item Using these approaches, we evaluate the strength of the sequential structure on 15 popular datasets from different domains. We identify the datasets that have weak sequential structure and are less appropriate for SRSs evaluation.
\end{itemize}

\section{Related Work}
\label{sec:related_work}
% recsys datasets
The area of the algorithms' evaluation improvement, including proper dataset selection, draws the increased attention of the research community \cite{Zhao_2022, Tamm_2021, meng2020exploring}.
A significant number of scientific papers employ a limited number of datasets, usually at most 3, to perform a model evaluation \cite{chin2022datasets}. The choice of datasets is often determined by the necessity of being aligned with the previous works to enhance reproducibility. Moreover, datasets are often chosen heuristically, which could influence the observations and/or conclusions obtained \cite{sun2020we, chin2022datasets}.
% We have examined the following works, which looked at ways to identify different patterns and characteristics in datasets.
% Datasets used vary greatly from paper to paper [some new links], and to the best of our knowledge, there is no reason why specific data sets should be used. In addition, datasets and a preprocessing for diverse recommendation tasks may differ. Therefore, researchers are increasingly thinking about issues of reproducibility and robustness of experiments (links).
Therefore, in the paper \cite{chin2022datasets}, a set of characteristics of datasets was proposed for their clustering and further selection of the most diverse datasets. An approach that would evaluate algorithms more robustly was also considered in the article \cite{shevchenko2024variability}, where various metrics aggregation methods were proposed to evaluate the quality of algorithms in general or on a specific domain.

% seq recsys datasets
The works above consider the top-K recommendation task, but the sequential recommendation task needs additional attention.
 % In the case of sequential recommendation algorithms, not all datasets are suitable for experiments. 
% Since the superiority of new algorithms over old ones significantly depends on the correct choice of datasets. 
In the case of SRSs evaluation, the data should contain sequential patterns that the algorithms can rely on \cite{hidasi2023widespread}.
% seq datasets analysis approaches
% Analysis of the dataset to identify user behavioral patterns and/or evaluate the datasets sequential nature was performed in several works. 
% Those approaches could be classified into model-free and model-based approaches, where the former are based on direct sequential pattern search and the latter utilize sequential model to indirectly evaluate the dataset characteristics.
In our work, we applied and extended several methods to evaluate the sequential structure of a wide range of datasets, frequently appearing in research papers dedicated to sequential recommendations. 

% model-free
An analysis of sequential patterns presence with sequential association rules \cite{han2022data} was performed by \cite{tang2018personalized} to prove the presence of union-level sequential patterns and skip behavior in real-world datasets. As the number of rules depends on a dataset's nature and is not directly comparable over multiple datasets, in our work, we propose comparing the number of rules before and after the users' sequences shuffling.
% The user preference drift is analyzed with the proportion of $i \to j$ item transitions by \cite{hidasi2023widespread}. 
% model-based

Analysis of sequential model performance on the original and shuffled sequences is another way to evaluate the sequential nature of the dataset, which was used by \cite{woolridge2021sequence} with the SASRec model.
The other close approach compares the performance of a sequential recommender and the same algorithm with the sequence modeling part replaced with a feedforward layer, shown in \cite{hidasi2023widespread}. 

We adopt the idea that the quality of a sequential recommender should decrease on shuffled data in the presence of sequential structure in a dataset. Unlike the work \cite{woolridge2021sequence}, we apply data shuffling to test sequences only and use the same trained model. Our approach allows to estimate the dataset's sequential patterns faster as we don't need to retrain the model.

Rank List Sensitivity approach with top-K Jaccard score calculation was applied by \cite{oh2022ranklist}. The authors proposed comparing two recommendation lists produced by the different versions of the recommender, trained on the same data, to quantify the model stability against perturbations. We adapt this approach to compare the recommendations obtained on shuffled and original sequences. This approach was also used in \cite{betello2023investigating} but for different data perturbations and goals.

% It is worth noting that in datasets using rating data, the presence of sequential patterns may be questionable because the time of rating is disjoint from the time of interacting with
% the item \cite{hidasi2023widespread}. So besides meeting formal criteria,  researchers need to pay attention to a dataset collection procedure.

%  The authors of the article \cite{woolridge2021sequence} proposed to use data shuffling to assess how much the quality of the algorithm will change, considered in the study SASRec \cite{kang2018self}. The authors of the article \cite{hidasi2023widespread} point out the importance of proper data preprocessing for comparing sequential recommendation algorithms.
% TODO: добавить про псевдо псевдо последовательные рекомендации

% In ML in general and in recommender systems in particular, the use of various datasets is necessary to confirm the superiority of the algorithm over previously created models. The vast majority of scientific articles use 3 or 4 datasets of different nature to confirm the results (links). As a rule, the choice of datasets is determined by the use of similar datasets in previous works. But at the same time, because The choice of datasets lies with the authors of the articles. Authors may also unknowingly select datasets that give their algorithm an advantage, or use datasets of the same type. Therefore, the issue of assessing the diversity and quality of datasets is one of the most important. The work (link) provides an analysis of the characteristics of datasets and their clustering for recommendation tasks.

\section{Methodology}
\label{sec:approach}

We apply several methods based on data shuffling to analyze sequential structure in datasets. We rely on the assumption that random shuffling of the order of interactions for each user will break sequential dependencies between items. The patterns identified in the shuffled sequences will appear here solely by chance. So, we can estimate the presence of sequential structure by measuring the discrepancy between the metrics acquired on shuffled and original data. If there are a lot of sequential patterns in the original dataset, the difference between the original and shuffled versions will be significant. The difference will be negligible if the original dataset has no sequential structure.

We suggest using two substantially different approaches to analyze datasets from various perspectives. The first one is simple and model-agnostic; it is based on  \textit{sequential rules identification}. The second one is model-based; it consists of \textit{training a sequential model and evaluating how model predictions change and deteriorate after shuffling}. The model-based approach is further divided into \textit{model performance deterioration} and \textit{change in top-K recommendation list}.

\subsection{Sequential rules}

We follow \cite{tang2018personalized} and mine sequential association rules with given support and confidence \cite{ han2022data}. Suppose we have sequences of user interactions $(i_1, i_2, ..., i_t)$. Then we count sequential rules of the form $(i_{t-L}, ..., i_{t-2}, i_{t-1}) \rightarrow i_t$, where the item $i_t$ follows directly after $(i_{t-L}, ..., i_{t-2}, i_{t-1})$ in the sequence. We use rules with orders $L=1$ and $L=2$, as some datasets have too few higher-order rules. So we count all item 2-grams $(i_j, i_k)$ and  3-grams $(i_j, i_k, i_l)$ that occur in the dataset. Support is computed as the total number of occurrences of these 2-grams and 3-grams. Confidence is computed as support of given n-gram divided by the total number of occurrences of $i_j$ for 2-gram and total number of occurrences of $(i_j, i_k)$ for 3-grams. In the end, we count the number of rules with minimum support and confidence greater than the chosen threshold values.

However, the number of such rules can not be directly compared across different datasets. It is highly dependent on the total number of users, number of items, sequence length, and item popularity. In order to normalize rule counts, we decide to explicitly destroy sequential patterns with shuffling and count the number of rules that occur by chance for a dataset with given characteristics. So, we can measure sequential structure by comparing results for original and shuffled versions of data, and this measure can be comparable across different datasets.

\subsection{Model-based approaches}

Model-based approaches involve training a sequential model, which is able to capture sequential patterns in data. The first way (our second approach) is to measure the model performance degradation in terms of recommendation accuracy metrics. Initially, we train and evaluate the model on the original dataset. Subsequently, we shuffle the test sequences, excluding holdout items, and evaluate the model on those shuffled sequences. We expect to observe the model performance deterioration if the dataset has strong sequential patterns. A similar approach was applied in \cite{woolridge2021sequence}, but shuffling was used not only for the test stage but also for the training stage.
% Our preliminary experiments with shuffling the training set yielded results similar to test sequences shuffling only. So, we use only the test set shuffling as we don’t need to retrain the model with this approach.

The second way (our third approach) is based on the Rank List Sensitivity approach with top-K Jaccard score \cite{oh2022ranklist}, which quantifies the model stability against perturbations by comparing two recommendation lists. We take the top-K recommendation list the model predicts for each user obtained with original and shuffled test sequences and compute mean Jaccard similarity \cite{jaccard1912distribution} between these lists. The Jaccard score measures the overlap in the top-K items without considering their order. The greater the overlap, the less the predictions differ; thus, the less the model relies on sequential patterns.

The results of the model-based approach can depend on the model used, as different models may highlight sequential properties in different ways. To overcome this limitation, we conduct experiments with two widely used sequential models of varying nature - SASRec \cite{kang2018self}, which is based on self-attention, and GRU4Rec \cite{hidasi2015session}, which is based on recurrent neural networks.

\begin{table}[!htbp]
\setlength{\abovecaptionskip}{2pt}
\setlength{\belowcaptionskip}{-7pt}
% \begin{minipage}{\columnwidth}
\caption{\textbf{Statistics of the datasets after preprocessing.}}
\resizebox{\columnwidth}{!}{%
\label{tab:datasetStats}
    \centering
    \begin{tabular}{llrrrrrr}
   \hline 
        \multirow{2}{*}{\textbf{Dataset}} & \multirow{2}{*}{\textbf{Domain}} & \multirow{2}{*}{\textbf{\# Users}} & \multirow{2}{*}{\textbf{\# Items}} & \multirow{2}{*}{\textbf{\# Interact.}} & \multicolumn{1}{c}{\textbf{Avg.}} & \multirow{2}{*}{\textbf{Density}} & \multirow{2}{*}{\textbf{\# Papers}} \\ 
        & & & & &\multicolumn{1}{c}{\textbf{length}} & &\\ \hline 
        
        \textbf{Beauty \cite{mcauley2015image}}  & \multirow{7}{*}{E-com} & 22,363 & 11,147 & 198,502 & 8.9 & 0.08\% & 45/35\% \\ 
        
        \textbf{Diginetica} 
        \footnote{\url{https://competitions.codalab.org/competitions/11161}} &  & 61,279 & 24,756 & 485,903 & 7.9 & 0.03\% & 19/15\%\\ 
        
        \textbf{OTTO \cite{normann2022ottodataset}}  &  & 122,233 & 127,395 & 2,552,310 & 20.9 & 0.02\% & \textbf{recent}\\ 
        
        \textbf{RetailRocket} \footnote{\url{https://www.kaggle.com/retailrocket/ecommerce-dataset}} &  & 38,582 & 20,228 & 463,992 & 12.0 & 0.06\% & 10/8\%\\ 
        
        \textbf{MegaMarket \cite{shevchenko2024variability}} &  & 186,189 & 411,767 & 12,672,359 & 68.1 & 0.02\% & \textbf{recent}\\ 
        
        \textbf{Sports  \cite{mcauley2015image}} &  & 35,598 & 17,342 & 296,337 & 8.3 & 0.05\% & 23/18\% \\ 
        
        \textbf{Yoochoose \cite{ben2015recsys}} &  & 1,557,906 & 19,063 & 13,009,746 & 8.4 & 0.04\% & 16/12\% \\
        
    \hline 
        \textbf{Games \cite{mcauley2015image}} & \multirow{2}{*}{Games} & 19,412 & 11,057 & 167,597 & 8.6 & 0.08\% & 35/27\%\\ 
        
        \textbf{Steam \cite{pathak2017generating}} &  & 281,210 & 7,931 & 3,484,694 & 12.4 & 0.16\% & 13/10\%\\ 
    \hline 
        \textbf{ML-20m \cite{harper2015movielens}} & Movies & 138,493 & 18,286 & 19,984,024 & 144.3 & 0.79\% & 14/11\% \\ 
    \hline 
        \textbf{30Music \cite{turrin201530music}} & \multirow{2}{*}{Music} & 43,762 & 822,507 & 22,604,876 & 516.5 & 0.06\% & 5/4\%\\
        
        \textbf{Zvuk \cite{shevchenko2024variability}} &  & 19,267 & 146,894 & 8,087,953 & 419.8 & 0.29\% & \textbf{recent} \\
    \hline 
        \textbf{Foursquare \cite{cheng2013you}} & \multirow{3}{*}{Soc. Net.} & 2,293 & 14,462 & 454,287 & 198.1 & 1.37\% & 6/5\%\\ 
        
        \textbf{Gowalla \cite{cheng2013you}} &  & 72,411 & 273,849 & 4,236,862 & 58.5 & 0.02\%  & 12/9\%\\ 
        
        \textbf{Yelp \cite{asghar2016yelp}} &  & 184,680 & 71,864 & 2,619,842 & 14.2 & 0.02\%  & 25/19\%\\ 
    \hline 
    
    \end{tabular}
    }
%     \bigskip
% \footnotesize\emph{1:} \url{https://competitions.codalab.org/competitions/11161}

% \emph{2:} \url{https://www.kaggle.com/retailrocket/ecommerce-dataset}
%\end{minipage}
\end{table}

\footnotetext[1]{\url{https://competitions.codalab.org/competitions/11161}}
\footnotetext[2]{\url{https://www.kaggle.com/retailrocket/ecommerce-dataset}}

\section{Experimental Settings}
\label{sec:experimental_settings}

\subsection{Datasets selection}

In order to identify the datasets widely used by the academic community for sequential recommenders evaluation, we performed an extended analysis of papers related to sequential recommendations from 3 top-tier conferences, RecSys, SIGIR, and CIKM, for the past five years (2019-2023). In total, we analyzed 130 papers and found that 98 public datasets have been used in one or more of these papers. We identified the most frequently used datasets and selected 12 datasets from different domains for our work. Table \ref{tab:datasetStats} contains the number and the percentage of papers that use each dataset. Further details of the analysis are included in the online appendix to our paper, which is available on GitHub.
% \footnote{\url{https://github.com/Antondfger/Does-It-Look-Sequential/}}.

In addition to commonly used academic benchmarks, we included some recently published industrial datasets, such as Zvuk, MegaMarket \cite{shevchenko2024variability}, and OTTO \cite{normann2022ottodataset}. To reduce computational costs while maintaining a sufficient amount of data for our analysis, we sampled 500,000 users of 12 million in the OTTO dataset.

\subsection{Datasets preprocessing}
As in previous publications \cite{tang2018personalized, kang2018self, sun2019bert4rec}, the presence of a review or rating is considered as implicit feedback. If the dataset contains multiple event types, we follow \cite{hidasi2023widespread} and leave only one event type corresponding to item views or clicks.

Following common practice \cite{kang2018self}, we discard users and items with a lower number of 
interactions. Namely, we applied N-core filtering with N equal to 5 for all datasets to preserve better reproducibility compared to the N-filter approach, where the results depend on filtering order \cite{sun2020we}. 
Following \cite{hidasi2023widespread}, we remove subsequent repeating items from the sequences, e.g., we change $(i, i, j)$ to $(i, j)$, but $(i, j, i)$  remains unchanged. 
The final statistics of the datasets after preprocessing are summarized in Table \ref{tab:datasetStats}.

\begin{figure}[htbp]
    \centering
    \setlength{\abovecaptionskip}{0pt}
    \includegraphics[width=\linewidth]{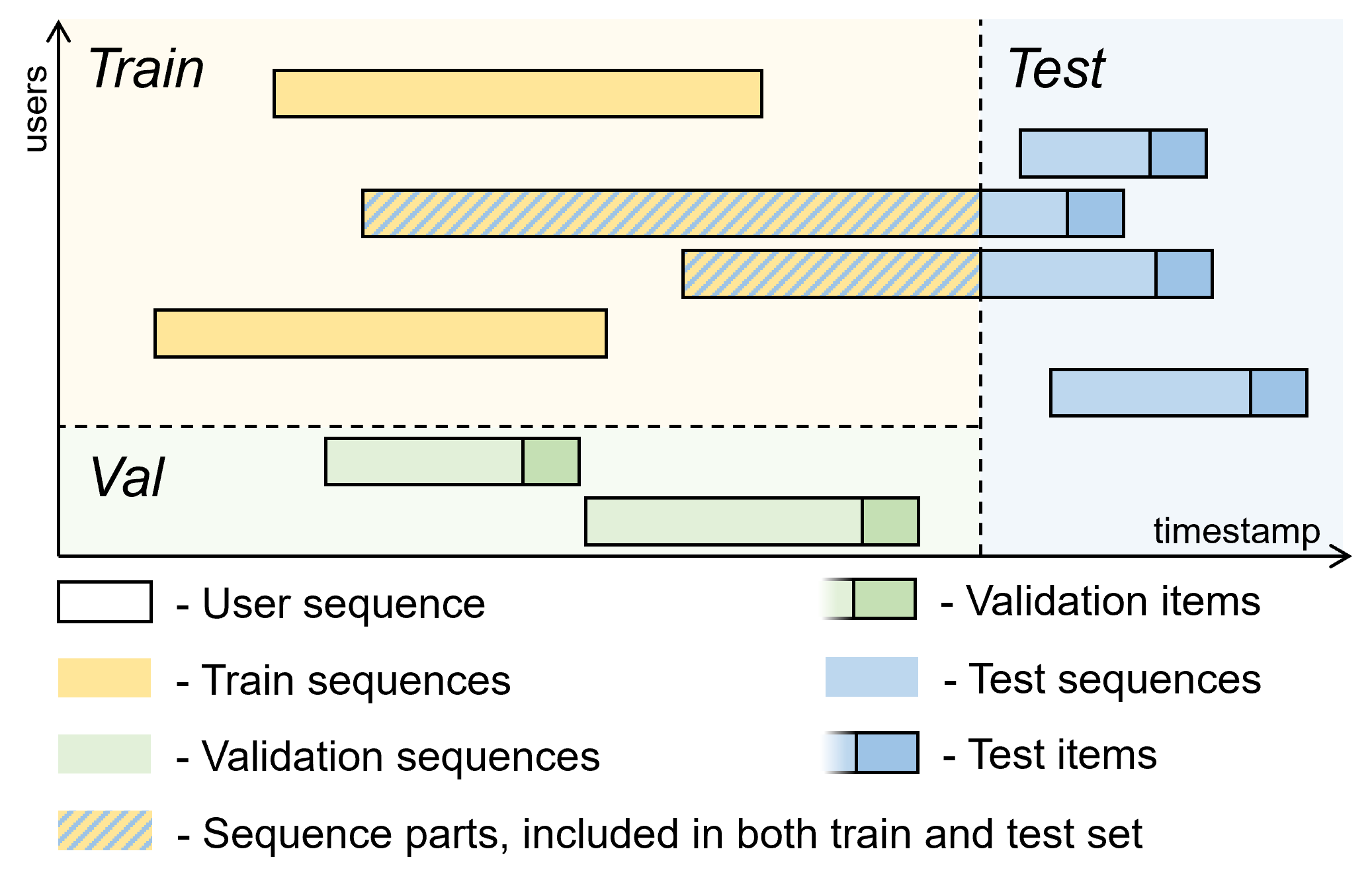}
    \caption{Data splitting strategy.}
    \Description[Splitting strategy]{Splitting strategy}
    \label{fig:split}
\end{figure}

\subsection{Evaluation}

One widely used evaluation strategy for sequential recommendations is the leave-one-out strategy applied by \cite{sun2019bert4rec, kang2018self}. The leave-one-out strategy is reasonable for the next-item prediction task, as it shows the model's sensitivity to sequential information and recent user preferences. Nevertheless, as recent publications showed, the leave-one-out strategy could lead to data leakage and may not truly reflect the performance of recommendation models in online settings \cite{ meng2020exploring, ji2023critical, sun2023take}.
In our work, we combine global temporal and leave-one-out split constraints to preserve the sequential nature of the recommendation task and avoid data leakage.

Thus, for each dataset, we select the global temporal boundary corresponding to 90\% of the interactions and take all preceding interactions into the training and validation subset, which are split randomly by the users. The validation set is used for early stopping. Each user interacting after the global temporal boundary is considered a test user. For the test and validation sets, we consider the last interaction of the users as the ground truth and all previous interactions as an input sequence. The splitting scheme is illustrated by Figure \ref{fig:split}. 

For performance evaluation, we use Normalized Discounted Cumulative Gain (NDCG@10) and HitRate@10 from the popular library Recommenders\footnote{\url{https://github.com/recommenders-team/recommenders}}. To compare recommendation lists, we use the top-K Jaccard score with $K=10$ (Jaccard@10).

\subsection{Implementation Details}

\begin{table*}%[!ht]
\setlength{\abovecaptionskip}{2pt}
\caption{Results for the model-based approaches with SASRec and GRU4rec models. Comparison of model performance before and after shuffling and Jaccard score between top-K recommendation lists. Relative changes greater than -10\% for accuracy metrics and Jaccard scores greater than $\frac{1}{3}$ indicate weak sequential structure and are marked in bold.}

\label{tab:model_based}
    \centering
    \resizebox{\textwidth}{!}{
    \begin{tabular}{?l?rrr|rrr?rrr|rrr?r|r?}
    \Xhline{3\arrayrulewidth}
        \multirow{4}{*}{\textbf{Dataset}}& 
        %------------------
        % \multicolumn{6}{c?}{\textbf{Sequential association rules}} &
        \multicolumn{6}{c?}{\textbf{SASRec performance degradation}} &   
        \multicolumn{6}{c?}{\textbf{GRU4Rec performance degradation}} &
        \multicolumn{2}{c?}{\textbf{Jaccard@10}} \\
        \cline{2-15}

        % & \multicolumn{3}{c|}{\textbf{2-grams}} & 
        % \multicolumn{3}{c?}{\textbf{3-grams}} 
        & 
        \multicolumn{3}{c|}{\textbf{HitRate@10}} & 
        \multicolumn{3}{c?}{\textbf{NDCG@10}} & 
        \multicolumn{3}{c|}{\textbf{HitRate@10}} &  
        \multicolumn{3}{c?}{\textbf{NDCG@10}} & 
        \multicolumn{1}{c|}{\multirow{3}{*}{\textbf{SASRec}}} &
        \multicolumn{1}{c?}{\multirow{3}{*}{\textbf{GRU4Rec}}} \\ 
        \cline{2-13} 
        
         % & \textbf{Before}  & \textbf{After}  & \textbf{Relative} & \textbf{Before}  & \textbf{After}  & \textbf{Relative} 
         & \textbf{Before}  & \textbf{After}  & \textbf{Relative} & \textbf{Before}  & \textbf{After}  & \textbf{Relative} & \textbf{Before}  & \textbf{After}  & \textbf{Relative}& \textbf{Before}  & \textbf{After}  & \textbf{Relative} & & \\ 
         % &  \textbf{shuffle} &  \textbf{shuffle} & \textbf{change} &  \textbf{shuffle} &  \textbf{shuffle}  & \textbf{change} 
         & \textbf{shuffle} &  \textbf{shuffle} & \textbf{change} &  \textbf{shuffle} &  \textbf{shuffle} & \textbf{change} &  \textbf{shuffle} &  \textbf{shuffle} & \textbf{change} &  \textbf{shuffle} &  \textbf{shuffle} & \textbf{change} & &\\

         \Xhline{3\arrayrulewidth}%\cline{2-21} 
         % & \textbf{B} & \textbf{A} & \textbf{R} & \textbf{B} & \textbf{A} & \textbf{R} & \textbf{B} & \textbf{A} & \textbf{R} & \textbf{B} & \textbf{A} & \textbf{R} & \textbf{B} & \textbf{A} & \textbf{R} & \textbf{B} & \textbf{A} & \textbf{R}  & \textbf{A}  & \textbf{A} 
         % \\ \Xhline{3\arrayrulewidth} 
          % &  \textbf{shuffle} &  \textbf{shuffle} & \textbf{change} &  \textbf{shuffle} &  \textbf{shuffle}  & \textbf{change} & \textbf{shuffle} &  \textbf{shuffle} & \textbf{change} &  \textbf{shuffle} &  \textbf{shuffle} & \textbf{change} &  \textbf{shuffle} &  \textbf{shuffle} & \textbf{change}
        \textbf{Beauty}  & 0.042 & 0.026 & -39\% & 0.019 & 0.011 & -43\% & 0.026 & 0.020 & -25\% & 0.013 & 0.010 & -26\% & 0.24 & 0.23 \\ 
        \textbf{Diginetica}  & 0.333 & 0.286 & -14\% & 0.161 & 0.149 & \textbf{-7\%} & 0.286 & 0.236 & -17\% & 0.145 & 0.122 & -16\% & \textbf{0.52} & \textbf{0.44} \\ 
        \textbf{OTTO}  & 0.205 & 0.143 & -30\% & 0.120 & 0.086 & -28\% & 0.158 & 0.095 & -40\% & 0.090 & 0.057 & -37\% & 0.28 & 0.13 \\ 
        \textbf{RetailRocket}  & 0.326 & 0.315 & \textbf{-4\%} & 0.195 & 0.190 & \textbf{-2\%} & 0.250 & 0.232 & \textbf{-7\%} & 0.143 & 0.135 & \textbf{-5\%} & \textbf{0.47} & \textbf{0.33}\\ 
        \textbf{MegaMarket}  & 0.192 & 0.101 & -47\% & 0.111 & 0.062 & -45\% & 0.184 & 0.078 & -58\% & 0.109 & 0.048 & -56\% & 0.19 & 0.10 \\ 
        \textbf{Sports} & 0.032 & 0.023 & -28\% & 0.016 & 0.011 & -32\% & 0.023 & 0.019 & -17\% & 0.012 & 0.010 & -18\% & 0.26 & \textbf{0.33}\\ 
        \textbf{Yoochoose}  & 0.396 & 0.308 & -22\% & 0.228 & 0.167 & -27\% & 0.384 & 0.286 & -26\% & 0.233 & 0.154 & -34\% & \textbf{0.46} & \textbf{0.38} \\ \hline
        \textbf{Games}   & 0.052 & 0.035 & -33\% & 0.025 & 0.015 & -38\% & 0.025 & 0.021 & -17\% & 0.012 & 0.010 & -17\% & 0.22 & 0.22\\ 
        \textbf{Steam}  & 0.110 & 0.099 & -10\% & 0.053 & 0.047 & -12\% & 0.109 & 0.101 & \textbf{-8\%} & 0.055 & 0.050 & \textbf{-9\%} & \textbf{0.59} & \textbf{0.56} \\ \hline
        \textbf{ML-20m}   & 0.075 & 0.031 & -59\% & 0.036 & 0.014 & -61\% & 0.082 & 0.031 & -63\% & 0.043 & 0.014 & -67\% & 0.12 & 0.07 \\ \hline
        \textbf{30Music}   & 0.198 & 0.020 & -90\% & 0.136 & 0.010 & -92\% & 0.204 & 0.011 & -95\% & 0.148 & 0.006 & -96\% & 0.12 & 0.02  \\ 
        \textbf{Zvuk} & 0.216 & 0.069 & -68\% & 0.112 & 0.034 & -70\% & 0.207 & 0.044 & -79\% & 0.121 & 0.022 & -82\% & 0.11 & 0.05\\ \hline
        \textbf{Foursquare}  & 0.353 & 0.328 & \textbf{-7\%} & 0.224 & 0.213 & \textbf{-5\%} & 0.108 & 0.108 & \textbf{0\%} & 0.060 & 0.061 & \textbf{0\%} & \textbf{0.39} & \textbf{1.00}\\ 
        \textbf{Gowalla}  & 0.301 & 0.277 & \textbf{-8\%} & 0.186 & 0.170 & \textbf{-8\%} & 0.225 & 0.215 & \textbf{-5\%} & 0.137 & 0.134 & \textbf{-2\%} & \textbf{0.45} & 0.26\\ 
        \textbf{Yelp}  & 0.044 & 0.043 & \textbf{-2\%} & 0.021 & 0.022 & \textbf{+5\%} & 0.037 & 0.034 & \textbf{-7\%} & 0.018 & 0.017 & \textbf{-7\%} & \textbf{0.37} & 0.3 \\ 
        \Xhline{3\arrayrulewidth}
    \end{tabular}
   }
\end{table*}

For the SASRec model, we use two self-attention blocks, two attention heads, and a hidden size 64. For the GRU4Rec model, we use one layer with a hidden size of 64. We set the maximum sequence length to 128 and batch size to 128. We train the models with cross-entropy loss as in \cite{klenitskiy2023turning} and use Adam optimizer with the learning rate 1e-3.

Following common practice \cite{tang2018personalized}, we apply support and confidence thresholds for sequential rules identification and analysis. We use a support threshold of 5 and a confidence threshold of 0.1. We also tried different threshold values and obtained similar results.

To increase the statistical significance of the results, we performed five experiment runs with different seeds for sequential rules mining and model-based approaches for each dataset. 

The code of our experiments is published on GitHub\footnote{\url{https://github.com/Antondfger/Does-It-Look-Sequential/}}.

\section{Results}
\label{sec:results}

\subsection{Sequential rules}
\begin{table}
\caption{Counts of sequential rules with chosen support and confidence threshold before and after shuffling. Shuffling has been performed 5 times and the average result is displayed. Relative changes greater than -90\% indicate weak sequential structure and are marked in bold.}
\label{tab:sequential_rules}
    \centering
    \resizebox{\columnwidth}{!}{
    \begin{tabular}{?l?rrr|rrr?}%rrr|rrr?rrr|rrr?r|r?}
    \Xhline{3\arrayrulewidth}
        \multirow{3}{*}{\textbf{Dataset}} 
        %------------------
        % \multicolumn{6}{c?}{\textbf{Sequential association rules}} \\
        %&  \multicolumn{6}{c?}{\textbf{SASRec performance degradation}} &   
        % \multicolumn{6}{c?}{\textbf{GRU4Rec performance degradation}} &
        % \multicolumn{2}{c?}{\textbf{Jaccard@10}} \\
        % \cline{2-7}

        & \multicolumn{3}{c|}{\textbf{2-grams}} & 
        \multicolumn{3}{c?}{\textbf{3-grams}} \\ 
        %& 
        % \multicolumn{3}{c|}{\textbf{Hit Rate@10}} & 
        % \multicolumn{3}{c?}{\textbf{NDCG@10}} & 
        % \multicolumn{3}{c|}{\textbf{Hit Rate@10}} &  
        % \multicolumn{3}{c?}{\textbf{NDCG@10}} & 
        % \multicolumn{1}{c|}{\multirow{3}{*}{\textbf{SASRec}}} &
        % \multicolumn{1}{c?}{\multirow{3}{*}{\textbf{GRU4Rec}}} \\ 
        \cline{2-7} 
        
         & \textbf{Before}  & \textbf{After}  & \textbf{Relative} & \textbf{Before}  & \textbf{After}  & \textbf{Relative} \\ 
         %& \textbf{Before}  & \textbf{After}  & \textbf{Relative} & \textbf{Before}  & \textbf{After}  & \textbf{Relative} & \textbf{Before}  & \textbf{After}  & \textbf{Relative}& \textbf{Before}  & \textbf{After}  & \textbf{Relative} & & \\ 
         &  \textbf{shuffle} &  \textbf{shuffle} & \textbf{change} &  \textbf{shuffle} &  \textbf{shuffle}  & \textbf{change} \\ 

         \Xhline{3\arrayrulewidth}%\cline{2-21} 
         % & \textbf{B} & \textbf{A} & \textbf{R} & \textbf{B} & \textbf{A} & \textbf{R} & \textbf{B} & \textbf{A} & \textbf{R} & \textbf{B} & \textbf{A} & \textbf{R} & \textbf{B} & \textbf{A} & \textbf{R} & \textbf{B} & \textbf{A} & \textbf{R}  & \textbf{A}  & \textbf{A} 
         % \\ \Xhline{3\arrayrulewidth} 
          % &  \textbf{shuffle} &  \textbf{shuffle} & \textbf{change} &  \textbf{shuffle} &  \textbf{shuffle}  & \textbf{change} & \textbf{shuffle} &  \textbf{shuffle} & \textbf{change} &  \textbf{shuffle} &  \textbf{shuffle} & \textbf{change} &  \textbf{shuffle} &  \textbf{shuffle} & \textbf{change}
        \textbf{Beauty} & 443 & 11.6 & -97\% & 131 & 0.0 & -100\% \\
        %& 0.042 & 0.026 & -39\% & 0.019 & 0.011 & -43\% & 0.026 & 0.020 & -25\% & 0.013 & 0.010 & -26\% & 0,24 & 0.23 \\ 
        \textbf{Diginetica} & 1464 & 379.4 & \textbf{-74\%} & 64 & 3.6 & -94\% \\%& 0.333 & 0.286 & -14\% & 0.161 & 0.149 & \textbf{-7\%} & 0.286 & 0.236 & -17\% & 0.145 & 0.122 & -16\% & \textbf{0,52} & \textbf{0.44} \\ 
        \textbf{OTTO} & 4907 & 491.8 & -90\% & 1942 & 69.0 & -96\% \\%& 0.205 & 0.143 & -30\% & 0.120 & 0.086 & -28\% & 0.158 & 0.095 & -40\% & 0.090 & 0.057 & -37\% & 0,28 & 0.13 \\ 
        \textbf{RetailRocket} & 2463 & 1145.0 & \textbf{-54\%} & 730 & 238.4 & \textbf{-67\%} \\%& 0.326 & 0.315 & \textbf{-4\%} & 0.195 & 0.190 & \textbf{-2\%} & 0.250 & 0.232 & \textbf{-7\%} & 0.143 & 0.135 & \textbf{-5\%} & \textbf{0,47} & \textbf{0.33}\\ 
        \textbf{MegaMarket} & 54265 & 1163.6 & -98\% & 35775 & 851.4 & -98\% \\%& 0.192 & 0.101 & -47\% & 0.111 & 0.062 & -45\% & 0.184 & 0.078 & -58\% & 0.109 & 0.048 & -56\% & 0,19 & 0.10 \\ 
        \textbf{Sports} & 129 & 7.4 & -94\% & 16 & 0.0 & -100\% \\%& 0.032 & 0.023 & -28\% & 0.016 & 0.011 & -32\% & 0.023 & 0.019 & -17\% & 0.012 & 0.010 & -18\% & 0,26 & \textbf{0.33}\\ 
        \textbf{Yoochoose} & 16848 & 3057.2 & \textbf{-82\%} & 73488 & 29528.2 & \textbf{-60\%} \\ \hline%& 0.396 & 0.308 & -22\% & 0.228 & 0.167 & -27\% & 0.384 & 0.286 & -26\% & 0.233 & 0.154 & -34\% & \textbf{0,46} & \textbf{0.38} \\ \hline
        \textbf{Games} & 207 & 17.4 & -92\% & 19 & 0.4 & -98\%  \\%& 0.052 & 0.035 & -33\% & 0.025 & 0.015 & -38\% & 0.025 & 0.021 & -17\% & 0.012 & 0.010 & -17\% & 0,22 & 0.22\\ 
        \textbf{Steam} & 247 & 1.2 & -100\% & 335 & 3.8 & -99\% \\ \hline%& 0.110 & 0.099 & -10\% & 0.053 & 0.047 & -12\% & 0.109 & 0.101 & \textbf{-8\%} & 0.055 & 0.050 & \textbf{-9\%} & \textbf{0,59} & \textbf{0.56} \\ \hline
        \textbf{ML-20m} & 536 & 0.0 & -100\% & 46727 & 4.8 & -100\%  \\ \hline%& 0.075 & 0.031 & -59\% & 0.036 & 0.014 & -61\% & 0.082 & 0.031 & -63\% & 0.043 & 0.014 & -67\% & 0,12 & 0.07 \\ \hline
        \textbf{30Music} & 273654 & 463.6 & -100\% & 267817 & 149.2 & -100\%  \\%& 0.198 & 0.020 & -90\% & 0.136 & 0.010 & -92\% & 0.204 & 0.011 & -95\% & 0.148 & 0.006 & -96\% & 0,12 & 0.02  \\ 
        \textbf{Zvuk} & 77209 & 414.0 & -99\% & 154009 & 172.4 & -100\% \\ \hline%& 0.216 & 0.069 & -68\% & 0.112 & 0.034 & -70\% & 0.207 & 0.044 & -79\% & 0.121 & 0.022 & -82\% & 0,11 & 0.05\\ \hline
        \textbf{Foursquare} & 3937 & 1641.0 & \textbf{-58\%} & 3628 & 797.4 & \textbf{-78\%} \\%& 0.353 & 0.328 & \textbf{-7\%} & 0.224 & 0.213 & \textbf{-5\%} & 0.108 & 0.108 & \textbf{0\%} & 0.060 & 0.061 & \textbf{0\%} & \textbf{0,39} & \textbf{1.00}\\ 
        \textbf{Gowalla} & 26181 & 11486.0 & \textbf{-56\%} & 9012 & 1635.0 & \textbf{-82\%} \\%& 0.301 & 0.277 & \textbf{-8\%} & 0.186 & 0.170 & \textbf{-8\%} & 0.225 & 0.215 & \textbf{-5\%} & 0.137 & 0.134 & \textbf{-2\%} & \textbf{0,45} & 0.26\\ 
        \textbf{Yelp} & 30 & 1.6 & -95\% & 4 & 0,0 & -100\% \\% & 0.044 & 0.043 & \textbf{-2\%} & 0.021 & 0.022 & \textbf{+5\%} & 0.037 & 0.034 & \textbf{-7\%} & 0.018 & 0.017 & \textbf{-7\%} & \textbf{0,37} & \textbf{0.3} \\ 
        \Xhline{3\arrayrulewidth}
    \end{tabular}
    }
\end{table}

\label{sec:sequential_rules_results}

Table \ref{tab:sequential_rules} contains results for the approach based on sequential rules. We can divide the datasets into two quite separate groups. For the first group, the relative change in the number of rules is less than -90\% and even close to -100\% for many of them. This indicates that these datasets might have fairly strong sequential patterns that break down after shuffling. However, this result should be taken with a grain of salt for some of these datasets (e.g., Beauty, Sports, Games, Steam, and especially Yelp) because the total number of rules was very small even before data shuffling.
For the second group, the relative change is significantly greater than -90\%, indicating that the datasets from this group might have a much weaker sequential structure. Diginetica, ReatailRocket, Yoochoose, Foursquare, and Gowalla are among these datasets.

This approach takes into account only a limited set of short-term sequential patterns, so we do not consider it a standalone measure of a dataset's sequential structure but rather an additional indicator.

\begin{figure*}[ht!]
    \centering
    \includegraphics[width=\textwidth]{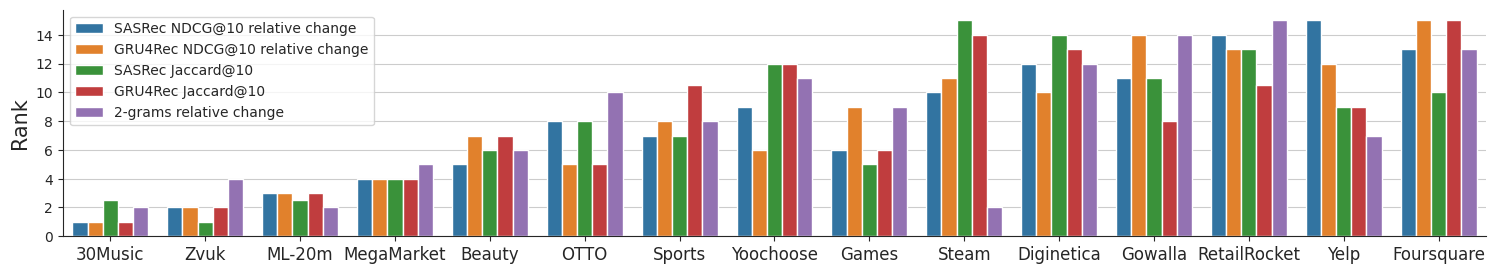}
    \caption{Ranks of the datasets according to different metrics considered. Rank 1 means this dataset has the strongest sequential structure according to the given metric. Datasets are sorted by an average of SASRec and GRU4Rec NDCG@10 ranks.}
    \Description[Ranks of the datasets according to different metrics considered]{Ranks of the datasets according to different metrics considered}
    \label{fig:results}
\end{figure*}
\subsection{Model-based approaches}

Results for the model-based approaches are shown in Table \ref{tab:model_based}. After shuffling, the performance drop in terms of NDCG@10 and HitRate@10 varies greatly depending on the dataset. The relative change for RetailRocket, Foursquare, Gowalla, and Yelp is very small for all metrics and models, indicating a weak sequential structure. Diginetica and Steam are borderline datasets with relatively weak sequential patterns as well. MegaMarket, ML-20m, 30Music, and Zvuk have a very strong sequential structure according to these metrics.

The top-K Jaccard score shows similar trends. Diginetica, RetailRocket, Youchoose, Steam, Foursquare, Gowalla, and Yelp have relatively high Jaccard similarity between recommendation lists for shuffled and original test sequences. The Jaccard@10 greater than $\frac{1}{3}$ indicates that the recommendation lists are similar for more than a half.   Data shuffling has a lower impact on the model outputs for these datasets, which is expected if the model hasn't learned to rely on sequential patterns. In contrast, the datasets with the most significant performance drop have a minimal Jaccard score. So, Jaccard@10 complements performance-based metrics and confirms the conclusions drawn from them.

\subsection{Influence of preprocessing}

It is worth noting that data preprocessing can influence the results of the analysis. For instance, some datasets contain numerous user sequences of short length. Sequential modeling may be of little importance for sequences of length 2 or 3.
In our main experiments, we use 5-core filtering, a popular preprocessing choice across many papers \cite{sun2020we}. However, we also conducted experiments for SASRec performance deterioration after filtering out only users with less than two interactions. While remaining stable for many datasets, the results have changed the most for Games and Sports. The relative change in NDCG@10 turned from -38\% to -6\% for Games and from -32\% to -10\% for Sports, indicating less pronounced sequential patterns. Thus, proper preprocessing may be required to analyze how well the algorithm captures the sequential structure.

\subsection{Comparison of approaches}

\begin{figure}[htbp]
    \centering
    \includegraphics[width=\linewidth]{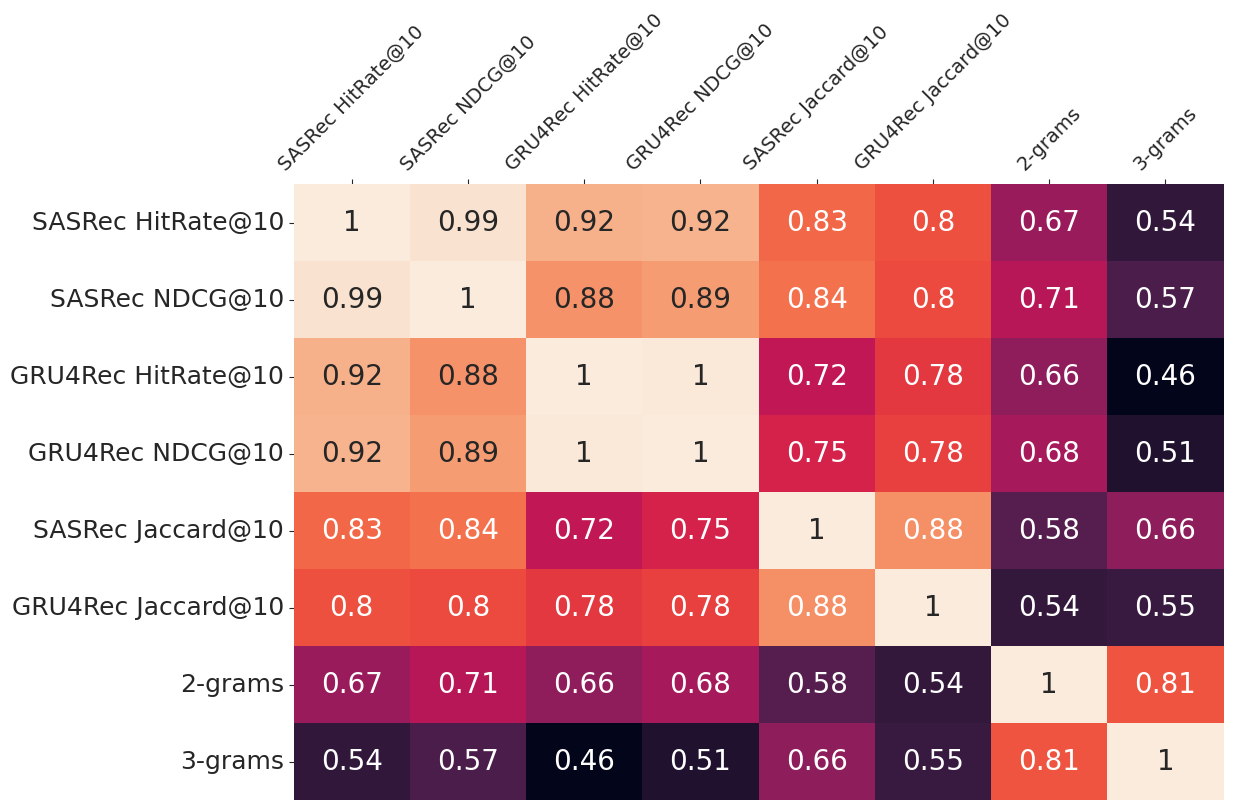}
    \caption{Spearman's correlation between all metrics: relative change in HitRate@10 and NDCG@10 for SASRec and GRU4Rec, Jaccard@10 for these models, and relative change in counts of 2-grams and 3-grams.}
    \Description[Spearman correlation between metrics.]{Spearman correlation between metrics.}
    \label{fig:corr_heatmap}
\end{figure}

Figure \ref{fig:corr_heatmap} shows Spearman's correlation coefficient between all metrics considered. In general, the correlation between different approaches is quite high. The correlation between relative changes in performance metrics for SASRec and GRU4Rec models is around 0.9. Both models show very similar results regarding sequential structure in the datasets. The approach based on sequential rules has a lower correlation with model-based approaches, which is expected. Relative change in the number of 3-grams has the lowest correlation. There are too few 3-grams even before shuffling for some datasets, and for 6 out of 15 datasets, the relative change is -100\%, making them indistinguishable. So, this is the least reliable metric.

In addition, Figure \ref{fig:results} illustrates the ranking of the datasets according to different approaches. In general, the results are quite consistent, although there is no perfect correspondence between all metrics. For example, the Youchoose dataset has significant sequential patterns according to model performance degradation but has bad results according to sequential rules count. However, this dataset has a low number of items and a very high number of interactions, so it can have a lot of 2-grams and 3-grams even after shuffling by chance. Unfortunately, our approach failed to capture this pattern. In contrast, Steam and Yelp are strong according to the model-agnostic approach and weak according to the model-based approach. But, as was stated in section \ref{sec:sequential_rules_results}, the total number of rules for these datasets was very small even before shuffling, indicating an absence of strong sequential structure.

% To summarize, Diginetica, RetailRocket, Steam, Foursquare, Gowalla, and Yelp should be considered as datasets with a weak sequential structure according to at least two of three approaches. Moreover, recommendation accuracy for all these datasets drops very slightly after shuffling the test data, which we consider to be the most important signal.
To summarize, we suggest that Foursquare,
Gowalla, RetailRocket, Steam, and Yelp are datasets with a weak sequential structure. Firstly, recommendation accuracy for all these datasets drops very slightly after shuffling the test data, which we consider to be the most important signal. Secondly, one or two additional approaches confirm this observation.

% \evspace{-10pt}
\section{Conclusion}
\label{sec:conclusion}

In this paper, we propose a set of three approaches to evaluating a dataset's sequential structure strength. We further analyze a wide range of datasets from different domains commonly used for the evaluation of SRSs. The results of our experiments show that many popular datasets, namely Foursquare, Gowalla, RetailRocket, Steam, and Yelp, lack a sequential structure.
Whether these datasets are suitable for evaluating sequential recommendations is questionable and needs further research.

The datasets selected for evaluation must be aligned with the task at hand. Conclusions drawn about the relative performance of different algorithms may change after selecting more appropriate datasets. Whether this is true or not is a possible future research direction, as well as further investigation of approaches to the assessment of sequential structure in datasets.

% The set of proposed approaches could be extended and improved with an application of more sophisticated models for sequential pattern identification, besides sequential rules, which we leave as a future research direction. 

To conclude, we encourage researchers to choose datasets for SRSs evaluation with careful consideration of their alignment with the sequential recommendation task.

\balance
\bibliographystyle{ACM-Reference-Format}
\bibliography{content/7_bibliography}
% \appendix 
% \section{Appendix}
% \label{appendix}

% \balance

\end{document}